\documentclass[twocolumn]{aastex631}

\usepackage{graphicx,color,amsmath,amssymb,bm}
\usepackage{slashed}
\usepackage{eqnarray}
\usepackage{orcidlink}

\newcommand{\nocontentsline}[3]{}
\newcommand{\toclesslab}[3]{\bgroup\let\addcontentsline=\nocontentsline#1{#2\label{#3}}\egroup}
\newcommand{\tocless}[2]{\bgroup\let\addcontentsline=\nocontentsline#1{#2}\egroup}

\usepackage{tikz}
\usepackage[normalem]{ulem}
\usepackage{tkz-euclide}
\usetikzlibrary{decorations.pathmorphing}	% To create gluons and vector bosons
\tikzset{
    v/.style={decorate, decoration={snake, segment length=3mm, amplitude=0.75mm}, draw},
    f/.style={draw=black, postaction={decorate},
        decoration={markings,mark=at position .6 with {\arrow[very thick]{latex}}}},
    fb/.style={draw=black, postaction={decorate},
        decoration={markings,mark=at position .4 with {\arrowreversed[very thick]{latex}}}},
    fnar/.style={draw=black},
    g/.style={decorate, draw=black,
        decoration={coil,amplitude=3pt, segment length=3.5pt}},
    s/.style={dashed,draw=black, postaction={decorate},
        decoration={markings,mark=at position .55 with {\arrow[very thick]{latex}}}},
    sb/.style={dashed,draw=black, postaction={decorate},
        decoration={markings,mark=at position .55 with {\arrowreversed[draw=black,very thick]{latex}}}},
    snar/.style={dashed,draw=black,line width =1.25pt},
}

\usepackage{amsmath}
\hypersetup{
    colorlinks=true,
    citecolor=blue,
    linkcolor=magenta,
    filecolor=magenta,
    urlcolor=blue,
}

\newcommand{\vecv}{\vec{v}}
\newcommand{\vecx}{\vec{x}}

\newcommand{\vecr}{\vec{r}}

\newcommand{\Msun}{\text{M}_\odot}

\newcommand{\msunkpc}{{\rm M}_\odot/{\rm kpc}^3}

\newcommand{\nablavec}{\vec{\nabla}}
\newcommand{\sigmaMW}{\sigma_{\scriptscriptstyle{\rm MW}}}

\newcommand{\sigmaSH}{\sigma_{\scriptscriptstyle{\rm SH}}}
\newcommand{\rhoSH}{\rho_{\scriptscriptstyle{\rm SH}}}
\newcommand{\rhoMW}{\rho_{\scriptscriptstyle{\rm MW}}}
\newcommand{\rsSH}{r_{\mathrm{s}, \scriptscriptstyle{\rm SH}}}
\newcommand{\vSH}{v_{\scriptscriptstyle{\rm SH}}}
\newcommand{\vecvSH}{\vec{v}_{\scriptscriptstyle{\rm SH}}}

\definecolor{mypurple}{RGB}{164,64,214}

\begin{document}
\raggedbottom

\title{Non-linear Evolution of Dark Plasma Subhalos}

\author{Andrew Liu, \orcidlink{0009-0008-0914-5130}}
\email{al1279@alumni.princeton.edu}
\affiliation{Department of Physics, Princeton University, Princeton, NJ 08544, USA}

\author{Anirudh Prabhu\,\orcidlink{0000-0001-9115-7844}}
\email{aprabhu@berkeley.edu}
\affiliation{Department of Physics, Princeton University, Princeton, NJ 08544, USA}
\affiliation{Leinweber Institute for Theoretical Physics, University of California, Berkeley, CA 94720, USA}
\affiliation{Theoretical Physics Group, Lawrence Berkeley National Laboratory, Berkeley, CA 94720, USA}

\author{Akaxia Cruz\,\orcidlink{0000-0001-7831-4892}}
\email{akaxia@princeton.edu}
\affiliation{Department of Physics, Princeton University, Princeton, NJ 08544, USA}
\affiliation{Department of Astrophysical Sciences, Princeton University, Princeton, NJ 08544, USA}
\affiliation{Center for Computational Astrophysics, Flatiron Institute, New York, NY 10010, USA}

\author{Mariangela Lisanti\,\orcidlink{0000-0002-8495-8659}}
\affiliation{Department of Physics, Princeton University, Princeton, NJ 08544, USA}
\affiliation{Center for Computational Astrophysics, Flatiron Institute, New York, NY 10010, USA}

\makeatletter
\let\frontmatter@title@above=\relax
\makeatother

\date{\today}

\begin{abstract} 
Dark matter that self interacts through long-range forces exhibits coherent, collective effects that are absent in short-range interactions. In the case where dark matter interacts through a hidden-photon mediator, its dynamics closely resemble those of Standard Model plasmas. In such models, various astrophysical environments, including cluster collisions and subhalos orbiting their host halo, are susceptible to plasma instabilities: processes that modify the dark matter velocity distribution and lead to exponential growth of dark electromagnetic fields. In this paper, we present the first study of the non-linear evolution of electrostatic instabilities in dark matter subhalos orbiting within the Milky Way potential using a suite of particle-in-cell simulations. We find that the growth and saturation of these instabilities produce substantial turbulent heating and mass loss, with an efficiency that depends sensitively on subhalo mass and orbital eccentricity. For highly eccentric orbits, plasma heating can reduce the initial mass of a $10^7$~($10^9$)~$\Msun$ subhalo by as much as $\sim$97\% ($\sim$84\%) soon after first pericenter. Plasma-induced heating and stripping may therefore leave observable signatures in the Milky Way subhalo population, including a suppression of the low-mass subhalo mass function.
\end{abstract}

% \maketitle

{\hypersetup{linkcolor=blue}
%%%%%%%%%%%%%% INTRODUCTION %%%%%%%%%%%%%%%%%%%%%%%%%%%%%
\noindent 

\section{Introduction}

Well-motivated alternatives to cold, collisionless dark matter~(DM) can modify galactic-scale structure~\citep{2018Natur.562...51B} while remaining consistent with large-scale astrophysical observations~\citep{Springel:2008cc, 2020}. Self-Interacting Dark Matter~(SIDM), for example, introduces a non-gravitational force confined to the dark sector and can affect the properties of individual dwarf galaxies and their population statistics---see reviews by \citet{Tulin:2017ara, Adhikari:2022sbh}. We focus here on the specific regime where the interaction range of this new force exceeds the typical interparticle spacing. Under these conditions, DM particles interact coherently with many neighbors simultaneously, driving collective---rather than binary---dynamics that closely resemble Standard Model plasmas.

The dark-$U(1)$ model provides a classic manifestation of this phenomenon.  In this scenario, the DM is composed of a dark electron and positron with charge $\mp q_\chi$ and mass $m_\chi$, and it self interacts via a dark photon of mass $m_{A'}$~\citep{HOLDOM1986196, Ackerman:2008kmp, Agrawal:2016quu}.  This model is generically compatible with early-universe observations, including from Big Bang Nucleosynthesis~\citep{Ackerman:2008kmp} and the linear matter power spectrum~\citep{Giffin:2025oqe}. It is further constrained by the Bullet Cluster, dwarf galaxy survival, and halo ellipticity, which rely on structure modification by dark Coulomb-like collisions~\citep{Feng:2009mn, Ackerman:2008kmp, Agrawal:2016quu}.

A distinctive feature of ``dark plasma'' models is that heat transport occurs through the scattering of particles off fluctuations in the collective mean field. As in Standard Model plasmas, large spatial and velocity-space gradients excite plasma instabilities (see, e.g.,~\citet{1973ppp..book.....K,Bhattacharjee2017}), which generate fluctuations in the system's background electromagnetic field. When these fluctuations reach sufficient amplitude, they scatter particles, modify their phase-space distribution, and drive the system toward a quasi-thermal equilibrium state~\citep{2023JPlPh..89e9016E, Banik:2024ijc, Ewart:2024bnh}. 

One canonical example is a streaming instability,\footnote{The two most common streaming instabilities are the ``two-stream'' and ``beam-plasma'' instabilities. Both involve two counter-propagating thermal plasmas. In the former case, the plasma densities are comparable; in the latter, they are very different.} in which two or more thermal plasmas move through one another with nonzero relative drift velocity.
Counter-streaming halos provide the necessary conditions to excite dark plasma streaming instabilities, which lead to the exponential growth of dark electrostatic and electromagnetic modes at the expense of the relative drift velocity~\citep{Lasenby:2020rlf, Cruz:2022otv, Medvedev:2024kjh}. Such instabilities are typically triggered when the relative drift velocity of the two DM halos is large compared to their internal velocity dispersions. Several recent studies have examined structural modifications driven by collective plasma instabilities~\citep{Lasenby:2020rlf, Cruz:2022otv, Medvedev:2024kjh, Giffin:2025oqe}, including numerical simulations of the Bullet Cluster~\citep{Heikinheimo:2015kra, Heikinheimo:2017meg, DeRocco:2024ifs}. 
In particular, \citet{DeRocco:2024ifs} studied the non-linear evolution of these instabilities using fully kinetic plasma simulations, setting leading constraints from the Bullet Cluster on the charge-to-mass ratio, $q_\chi/m_\chi < 2 \times 10^{-14} \ {\rm GeV}^{-1}$, which is orders-of-magnitude stronger than collisional constraints from halo ellipticity when the dark photon is sufficiently light~\citep{Ackerman:2008kmp, Feng:2009mn, Agrawal:2016quu}.} 

In this work, we instead focus on the infall of DM subhalos into the Milky Way~(MW). Within the subhalo's virial radius, the counter-streaming subhalo and MW plasmas create a velocity-space configuration that can also excite streaming instabilities. To capture these dynamics, we present the first fully kinetic, non-linear study of subhalo evolution using a suite of one-dimensional particle-in-cell~(PIC) simulations with the TRISTAN-MP code~\citep{tristan-mp-pu}. We demonstrate how the growth and saturation of electrostatic waves heat and evaporate DM particles, driving an effective stripping of weakly bound material. By quantifying this mass loss across different subhalo properties and orbital parameters, we provide an initial estimate of how dark electromagnetic interactions may reshape the MW subhalo mass function.

The paper is organized as follows. Section~\ref{sec:basics} introduces the basic scaling relations needed to understand the behavior of a dark plasma halo.  Then, Section~\ref{sec:methodology} outlines the overall methodology used to study the evolution of subhalos in the MW.  Results of the non-linear plasma evolution and its impact on subhalo dynamics are presented in Section~\ref{sec:results}.  Section~\ref{sec:discussion} describes potential extensions to the modeling, and Section~\ref{sec:conclusion} concludes.  Additional contextual and technical details are provided in the Supplemental Material. 

\section{Basics of Dark Plasma Halos}
\label{sec:basics}

We model an isolated, spherical DM halo as a net-neutral plasma of particles with mass $m_\chi$ and charge $\pm q_\chi$, interacting through gravity and a dark electromagnetic force. The latter is mediated by a dark photon of mass $m_{A'}$.  The halo's phase-space distribution $f_\pm(\vecx,\vecv)$ evolves according to the Boltzmann equation,
\begin{align}\label{eqn:vlasov}
    {\partial f_\pm \over \partial t} + \vec{v} \cdot \nablavec f_\pm - \nablavec \left[\Phi_{\rm g} \pm {q_\chi \Phi_{\rm em} \over m_\chi} \right]&  \cdot \nablavec_v f_\pm  \nonumber \\
    &= C_\pm[f_+, f_-] \,,
\end{align}
where $f_+$ and $f_-$ are the dark positron and electron distribution functions, respectively, and $\Phi_{\rm g}$ and $\Phi_{\rm em}$ are the gravitational and dark electrostatic potentials, which satisfy their respective Poisson equations,
\begin{equation}\label{eqn:poisson-g}
    \nabla^2 \Phi_{\rm g} = 4\pi G m_\chi \displaystyle\int d^3 v \, [f_- + f_+] \nonumber
\end{equation}
and 
\begin{equation}\label{eqn:poisson-em}
    \nabla^2 \Phi_{\rm em} = q_\chi \displaystyle\int d^3 v \, [f_- - f_+] \,.
\end{equation}
For simplicity, we neglect the Lorentz force from the magnetic field and return to this issue in Section~\ref{sec:discussion}. The collision operator, $C_\pm[f_+, f_-]$, accounts for interparticle interactions. In a plasma, Coulomb interactions drive the particle distribution toward thermal equilibrium on the collisional relaxation timescale~\citep{2002ctmp.book.....H}. Because Coulomb scattering is dominated by many small-angle encounters rather than isolated large-angle collisions, the relevant timescale is set by the time required for cumulative scattering to produce an order-unity change in a particle's velocity. This corresponds to a collision frequency, 
\begin{align} \label{eqn:nu_coll}
    \nu_{\rm coll} \simeq & \frac{1}{10 \, {\rm Gyr}} \, \left( \frac{\ln\Lambda}{100}\right) \, \left( \frac{q_\chi}{3 \times 10^{-3} \, e} \right)^4 \left( \frac{m_\chi}{m_p} \right)^{-3} \nonumber \\
    &\hspace{0.2in}\times \left( \frac{\rho}{10^7 \ \msunkpc} \right) \left( \frac{\sigma}{100 \, {\rm km/s}}\right)^{-3}  \, ,\end{align}
where $\ln\Lambda$ is a Coulomb logarithm factor that is defined below, $e$ and $m_p$ are the proton charge and mass, respectively, $\rho$ is the mass density of the dark plasma, and $\sigma^2/2 = \left\langle v_\chi^2 \right\rangle/3$ with $\left\langle v_\chi^2 \right\rangle$ the root-mean-square~(RMS) velocity of the particles. When the relaxation time, determined by the inverse collision frequency in Eq.~\eqref{eqn:nu_coll}, is much larger than the age of the Universe, we can safely neglect the collision operator in Eq.~\eqref{eqn:vlasov}.

In contrast, collective, kinetic effects take place on a much smaller timescale. This is set by the plasma frequency, which quantifies the response of a homogeneous plasma to perturbations and is defined by
\begin{align}\label{eqn:plasma_frequency}
    \omega_{\rm p}  \approx {1 \over 4 \, {\rm kyr}} \left( \frac{q_\chi/m_\chi}{10^{-14} \, e/m_p} \right) \left( \frac{\rho}{10^7 \ \msunkpc} \right)^{1/2}.
\end{align}
If a fluid element is displaced by a distance $\delta x$, an electric field is generated such that the equation of motion of the fluid element is $\ddot{x} = - \omega_{\rm p}^2 x$. As evident from Eq.~\eqref{eqn:plasma_frequency}, the plasma oscillation timescale can be comparable to galactic timescales for a charge-to-mass ratio more than fourteen orders of magnitude smaller than that of the proton, for characteristic densities. Throughout this work, we restrict our attention to this collisionless regime and adopt $q_\chi/m_\chi = 10^{-14} e/m_p \approx 3 \times 10^{-15} \, {\rm GeV}^{-1}$ as a benchmark. We will motivate this benchmark choice in the following sections.

A related scale is the Debye length,
\begin{align}
    \label{eq:debye}
    \lambda_{\scriptscriptstyle{\rm D}} &\approx 0.4 \, {\rm pc} \left(\sigma \over 100 \, {\rm km/s} \right) \left( \frac{q_\chi/m_\chi}{10^{-14} \, e/m_p} \right)^{-1} \nonumber\\  
    & \hspace{0.9in} \times \left( \frac{\rho}{10^7 \ \msunkpc} \right)^{-1/2} \, ,
\end{align}
which characterizes the distance at which a charge is screened in the plasma. To treat a halo of virial radius $r_{\scriptscriptstyle{\rm vir}}$ as a continuous, smooth medium, the condition $\lambda_{\scriptscriptstyle{\rm D}}/r_{\scriptscriptstyle{\rm vir}} \ll 1$ must be satisfied, which is indeed the case for the systems considered in this work. Additionally, the dark photon mass must satisfy 
\begin{align} \label{eqn:dark_photon_mass}
    m_{A'} &\lesssim \lambda_{\scriptscriptstyle{\rm D}}^{-1} \approx 2 \times 10^{-23} \, {\rm eV} \left( \frac{q_\chi/m_\chi}{10^{-14} \, e/m_p} \right)  \nonumber \\
    & \hspace{0.5in} \times \left(\sigma \over 100 \, {\rm km/s} \right)^{-1} \left( \frac{\rho}{10^7 \ \msunkpc} \right)^{1/2}
\end{align}
to ensure that the dark electrostatic force extends beyond the Debye length and collective effects arise.

Moreover, for the parameter space of interest, the number of particles per Debye sphere, defined as $\Lambda \equiv \frac{4\pi}{3} \left(\rho/m_\chi \right) \lambda_{\scriptscriptstyle{\rm D}}^3$, is significantly greater than unity. The plasma parameter, $\Lambda$, determines the relative importance of binary collisions versus collective electromagnetic interactions. When $\Lambda \gg 1$, it shows up in the Coulomb logarithm factor in Eq.~\eqref{eqn:nu_coll}. In the limit $\Lambda \gg 1$, many particles occupy a Debye volume and the fields experienced by individual particles arise from the superposition of many weak contributions. The plasma is then weakly coupled, with potential energy per particle small compared to kinetic energy, and is well-described by a mean-field (Vlasov) framework in which collective modes and screening govern the system. 

We consider regions of parameter space in which $\Lambda \ggg 1$, so that the dynamics are, to an excellent approximation, collisionless. In this regime, other radiative processes such as annihilation and Bremsstrahlung cooling are likewise strongly suppressed. The annihilation cross section scales as $\sigma_{\rm ann} \sim q_\chi^4/(m_\chi^2 \sigma)$, which is parametrically suppressed compared to the scattering cross section by a factor of $\sigma^3$, making annihilation subdominant to the relaxation time, $\nu_{\rm coll}^{-1}$. Radiative cooling processes, such as dark Bremsstrahlung, are further suppressed by additional powers of the coupling, with rates proportional to higher powers of $q_\chi$ (e.g., $\propto q_\chi^6$ up to kinematic and logarithmic factors). As a result, both particle number-changing and energy-loss processes are inefficient. Thus, the plasma evolution is dominated by collective, essentially collisionless dynamics.

\section{Semi-Analytic Model} \label{sec:methodology}

This section outlines the methodology used to model the evolution of a DM subhalo within the MW host. Section~\ref{sec:equilibrium} describes the equilibrium plasma state of the combined subhalo–MW system and introduces its initial configuration. Secs.~\ref{sec:orbit_sims} and \ref{sec:simulations} then review the numerical procedure used to follow the evolution of the subhalo deep into the nonlinear regime.  The semi-analytic framework described here enables a first comparison of the subhalo mass loss expected from plasma heating to that from standard tidal stripping.

\subsection{Initializing the Halos} \label{sec:equilibrium}

Following Eq.~\eqref{eqn:vlasov}, we adopt a simple class of steady-state ($\partial_t f_\pm = 0$) equilibrium solutions by enforcing local charge neutrality, $f_+ = f_-$. In this limit, the dark electrostatic potential vanishes, and the system reduces to a standard collisionless gravitational equilibrium with distribution function $f = f_+ + f_-$, where $f$ corresponds to an equilibrium CDM solution. In dark plasmas, however, such CDM-like equilibria (e.g., triaxial configurations) may be unstable to the generation of macroscopic electromagnetic fields, potentially leading to substantial modifications of halo structure. A broader discussion of equilibrium solutions and their stability properties will be presented in future work~(\citeauthor{otherpaper}, \textit{in preparation}). For simplicity, we restrict our attention to CDM-like initial halo configurations throughout this paper.

We approximate the DM velocity distribution of halo $i \in \{{\rm subhalo}, {\rm MW} \}$ as an isotropic Maxwell-Boltzmann, 
\begin{equation}
    f_i(\vecr, \vecv) = \frac{\rho_i(r)}{m_\chi} 
{\exp\left[- v^2 / \sigma_i^2(r)\right] \over (\pi\sigma_i^2(r))^{3/2}} \, ,
\end{equation}
where $\rho_i(r)$ is the spherically symmetric density distribution of the halo, and $\sigma_i(r)$ is its associated velocity dispersion distribution. The density\footnote{This does not yield a self-consistent solution to the static, isotropic Jeans equation. For a spherical density distribution, Eddington's  formula~\citep{Eddington1916} can be used to obtain the corresponding velocity distribution, which need not be locally Maxwell-Boltzmann. A detailed treatment of these effects is left to future work.} of halo $i$ is initialized to a Navarro-Frenk-White~(NFW) profile, 
\begin{equation}
\rho_i(r) = \frac{\rho_{\mathrm{s}, i}}{(r/r_{\mathrm{s},i})(1 + r/r_{\mathrm{s}, i})^{2}}
\end{equation}
with scale density $\rho_{\mathrm{s}, i}$ and scale radius $r_{\mathrm{s}, i}$~\citep{Navarro:1996gj}.  For the MW host, we assume a mass of $1.2 \times 10^{12}~{\rm M}_\odot$ and concentration\footnote{The halo concentration is defined as $c_i \equiv r_{\mathrm{vir},i}/r_{\mathrm{s},i}$, where the virial radius $r_{\mathrm{vir},i}$ is defined as the radius enclosing a mean density of $200\,\rho_{\rm crit}$, and $\rho_{\rm crit}$ is the critical density of the Universe, computed using the cosmological parameters in~\citet{2020}.} of $c_{\scriptscriptstyle{\rm MW}} = 8.2$~\citep{Dutton_2014}.  For the subhalo, we consider two specific cases for the virial mass: $10^7$ and $10^9~{\rm M}_\odot$ with associated concentrations $c_{\scriptscriptstyle{\rm SH}} = 27$ and 17, respectively~\citep{Dutton_2014}. While we adopt a CDM concentration-mass relation as a starting point, this relation may be modified in the dark plasma framework, particularly if mergers induce additional heating and mass loss.

At a given radius, the velocity dispersion of a halo can be estimated as $\sigma_i^2(r) \simeq G M_{{\rm enc},i}(r)/r$, where $G$ is the gravitational constant, and $M_{{\rm enc},i}$ is the enclosed mass.  For simplicity, we approximate the halo dispersion as constant and roughly equal to its value at the scale radius, $\sigma_i^2 \simeq G \rho_{\mathrm{s},i}^{} r_{\mathrm{s},i}^2$. For a NFW profile, the velocity dispersion peaks near the scale radius. Approximating it as uniform therefore overestimates the dispersion away from this radius. Since higher dispersions suppress plasma heating and enhance tidal stripping, this choice provides a conservative comparison between the two evaporation mechanisms. Using this approximation, the initial velocity dispersions are $\sigmaMW = 120$~km/s for the MW and $\sigmaSH = 2$ and $10$~km/s for the $10^7$ and $10^9~{\rm M}_\odot$ subhalo, respectively.

Taking the position and velocity of the subhalo relative to the host to be $\vec{r}_{\scriptscriptstyle{\rm SH}}$ and $\vec{v}_{\scriptscriptstyle{\rm SH}}$, respectively, we approximate the total equilibrium distribution function of the system as 
\begin{equation}
    f(\vecr, \vecv) = f_{\scriptscriptstyle{\rm MW}}(\vecr, \vecv) + f_{\scriptscriptstyle{\rm SH}}(\vecr - \vec{r}_{\scriptscriptstyle{\rm SH}}, \vecv - \vec{v}_{\scriptscriptstyle{\rm SH}})  \, .
    \label{eq:equilibriumf}
\end{equation} 
Equilibrium solutions of this form with multiple peaks are susceptible to two-stream or beam-plasma (depending on the subhalo-to-MW density ratio) instabilities---see App.~\ref{app:linear_stability} for a review. If such an instability occurs, it can result in localized heating that enhances the subhalo's evaporative mass loss. The remainder of this section describes how these effects are integrated into the semi-analytic model.

\subsection{Modeling Subhalo Orbits and Mass Loss}
\label{sec:orbit_sims}

\begin{figure*}
    \centering
    \includegraphics[width=\linewidth]{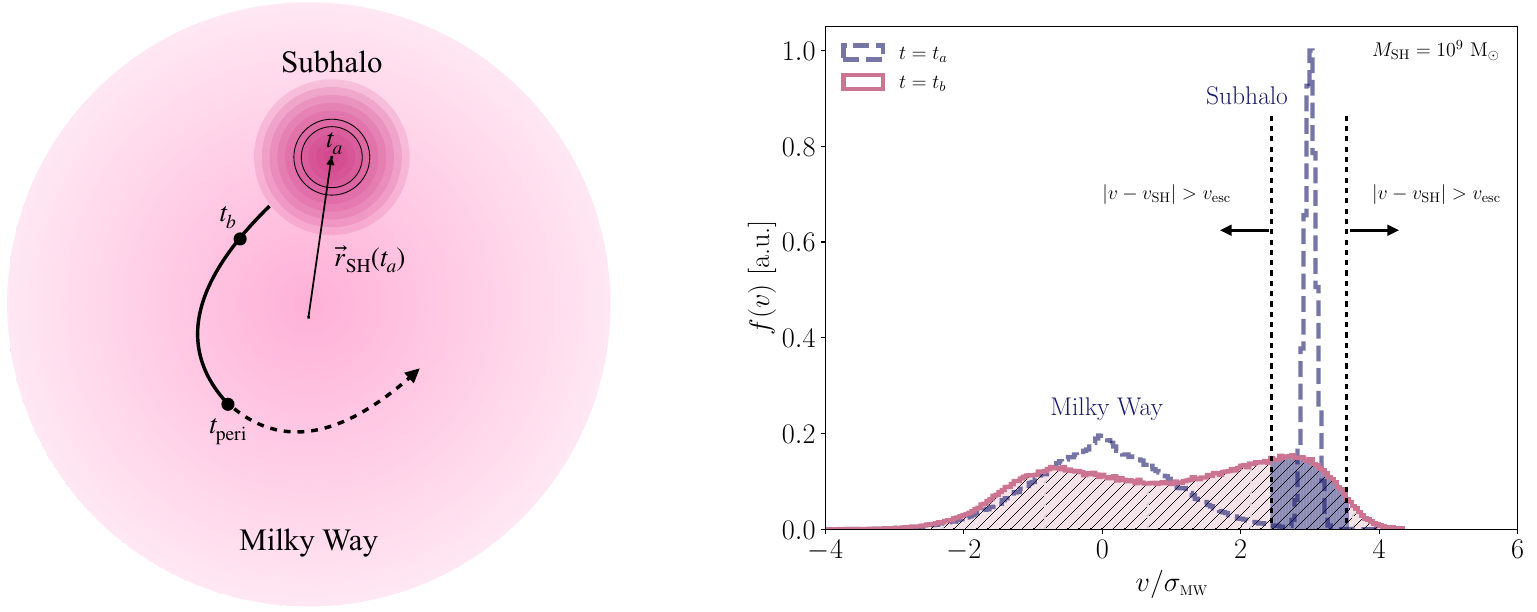}
    \caption{Schematic overview of subhalo evolution and phase-space structure in the host halo. (\emph{Left})~A subhalo on an eccentric orbit within the MW host. The orbit is marked at representative times $t_a$, $t_b > t_a$, and $t_{\rm peri}$, near the first orbital pericenter. The subhalo is sub-divided into radial slices with the $n^{\rm th}$ slice located at radius $r_n$. At time $t_a$, the position of the subhalo center-of-mass is $\vec{r}_{\scriptscriptstyle{\rm SH}}(t_a)$. (\emph{Right})~One-dimensional velocity distribution $f(v)$ within the radial slice $n$ for a representative subhalo ($M_{\scriptscriptstyle{\rm SH}} = 10^9 \ {\rm M}_\odot$), shown at two times ($t=t_a$ and $t=t_b$). The corresponding phase-space parameters are $\rhoSH(r_n)/\rhoMW(r_{\scriptscriptstyle{\rm SH}}(t_a)) = 1/2$, $\sigmaSH/\sigmaMW = 0.1$, and $\vSH/\sigmaMW = 3$. The narrow peak at large $v/\sigmaMW$ corresponds to the subhalo component, while the broader distribution near $v \sim 0$ arises from the MW background. The vertical dashed lines mark the escape condition $|v - \vSH| > v_{\rm esc} \approx 4 \sigmaSH$, separating bound from unbound particles. The evolution from $t_a$ to $t_b$ illustrates the progressive stripping of material and the redistribution of particles in phase space.}
    \label{fig:subhalo_MW}
\end{figure*}

We use \texttt{galpy}~\citep{Bovy_2015} to model the subhalo orbit.  To simplify the orbital integration, the subhalo is approximated as a point particle evolving within the MW gravitational potential, described by the \texttt{MWPotential2014} model. The potential consists of a disk described by a Miyamoto-Nagai potential~\citep{1975PASJ...27..533M}, a central bulge described by a power-law density profile with an exponential cutoff, and a DM halo described by a NFW potential (which we modify to be consistent with the parameters given in Section~\ref{sec:equilibrium}). The subhalo is initialized at the virial radius of the host ($\sim 220$~kpc) in the galactic plane. Initial tangential and radial velocities are chosen to produce orbits of eccentricities $\epsilon = 0.1, 0.3, 0.5, 0.7, 0.9$. The total speed of a subhalo at infall is the virial speed of the MW host, $\approx 140$~km/s. This procedure provides the location of the subhalo as a function of time, $\vec{r}_{\scriptscriptstyle{\rm SH}}(t)$. 

The goal is to model the mass loss of the subhalo as it evolves along its orbit. {To approximate this, we divide the subhalo density profile into a series of concentric radial shells using linear binning out to $r_{\rm max}=r_{\scriptscriptstyle{\rm vir,SH}}$, where $r_{\scriptscriptstyle{\rm vir,SH}}$ is the subhalo virial radius. We then restrict our analysis to radii $r\geq r_{\rm min}=0.1\rsSH$, where $\rsSH$ is the subhalo scale radius, because at smaller radii the subhalo velocity dispersion decreases rapidly and our PIC simulation results are no longer reliable.}
The number of bins is set to $3 c_{\scriptscriptstyle{\rm SH}}$, resulting in a bin width of $r_{\scriptscriptstyle{\rm s,SH}} / 3$. We assume that the $n^{\rm th}$ shell has a uniform density, $\rhoSH(r_n)$, that  corresponds to the volume-averaged value of the NFW profile between the shell's inner and outer radius.  

{To track how the density of each shell evolves with time, we discretize the orbit in time steps with $\delta t = 7$~Myr increments, cutting off the evolution of the density profiles at $t = 1.2 \, t_{\rm peri}$ for each of the orbits, where $t_{\rm peri}$ is the time of first pericenter.}  The spacing $\delta t$ is chosen to be much smaller than the subhalo’s orbital period and much greater than the  characteristic timescale for the saturation of electrostatic plasma instabilities (see Eq.~\eqref{eqn:plasma_frequency}). The latter condition ensures that dark plasma instabilities saturate within each timestep of the orbit, because the time to do so is much shorter than the dynamical time by construction. 

The left panel of Figure~\ref{fig:subhalo_MW} provides an illustration of the basic setup, with the subhalo at position $\vec{r}_{\scriptscriptstyle{\rm SH}}(t_a)$ along its orbit about the MW. The mass loss from plasma heating and tidal stripping is evaluated in each radial slice at a given point in time. The right panel shows the velocity-space distribution function within a representative radial slice of the subhalo centered at radius $r_n$. The slice is assumed to be sufficiently thin compared to the subhalo scale radius that the distribution function may be treated as spatially uniform. The example shown corresponds to a subhalo of mass $M_{\rm SH}=10^9 \ {\rm M}_\odot$, with local density contrast $\rhoSH(r_n)/\rhoMW(r_{\scriptscriptstyle{\rm SH}}(t_a)) = 1/2$ and streaming speed $\vSH=3\sigmaMW$. The initial distribution at time $t_a$ is shown in blue, while the distribution at a later time $t_b$, after the instability has saturated, is shown in pink. The shaded blue region denotes particles whose velocities, in the subhalo rest frame, lie below the local escape velocity. These particles remain gravitationally bound to the subhalo, whereas particles outside this region become unbound.  

At each time step, we determine how tidal stripping and plasma heating independently alter the density of each shell, treating these effects separately to compare their relative contributions. To account for tidal effects, we adopt the model used in \citet{Green_2021}, where mass is stripped from a subhalo outside of an instantaneous tidal radius, $r_{\rm tid}$, determined at each point within its orbit. At a given timestep, the tidal radius is computed and the shells outside this radius ($r_n > r_{\rm tid}$) evolve as
\begin{align}
    \rhoSH (r_n > r_{\rm tid},  t_{j+1}) = \left[1 - \alpha\, \frac{\delta t}{t_{\rm dyn}}\right]  \, \rhoSH(r_n > r_{\rm tid}, t_j) \, ,
\end{align}
where $\alpha = 0.55$ is a tidal stripping efficiency parameter determined from idealized simulations \citep{Green_2021, Jiang_2021} and $t_{\rm dyn}$ is the orbital dynamical time.

Modeling the mass loss due to plasma heating requires plasma PIC simulations, which are described in the next subsection.  Standard configurations for PIC simulations initialize two counter-streaming beams of plasma with uniform density in a box with periodic boundary conditions. The use of periodic boundary conditions effectively models the beams as infinite, which is sufficient to study the local effects of plasma instabilities (e.g., within a density shell of the subhalo). In the case of the subhalo-MW system, when viewing the system in the subhalo’s rest frame, background particles from the MW are constantly entering and exiting the subhalo as it moves through the ambient plasma, which makes these periodic boundary conditions sufficient to describe the system. In the PIC configuration, the relative beam density is reflective of the density ratio between the subhalo's DM and that of the background MW at a specific spatial location, while the beam velocity is reflective of the subhalo velocity along its orbit.  Specifying these two quantities provides the DM phase-space distribution, $f(\vecr, \vecv)$, as a function of time. For a spatial point located within the subhalo's virial radius, one can then calculate the quantity, $f_{\rm bound}$, which corresponds to the fraction of particles with speed less than $4\sigmaSH$,  a conservative estimate for the subhalo escape velocity.\footnote{For a NFW distribution, the ratio of the escape velocity to the local velocity dispersion diverges as $r\to 0$. In our PIC simulations, however, we assume a constant subhalo velocity dispersion, $\sigmaSH=\sigmaSH(\rsSH)$. Under this approximation, $v_{\rm esc}(r)/\sigmaSH$ approaches $3.3$ as $r\to 0$ and decreases with increasing radius. Our adopted criterion, $v_{\rm esc}=4\sigmaSH$, therefore overestimates the escape velocity throughout the subhalo, classifying more particles as bound and yielding a conservative estimate of the mass loss.} 

{The density of a shell in the subhalo is determined iteratively following 
\begin{align} \label{eqn:rho_stripping}
    \rhoSH (r_n, t_{j+1}) = \left[1 - {\frac{(1 - f_{\rm bound}(r_n, t_j)) \, \delta t}{\tau_{\scriptscriptstyle{\rm SH}}}}\right] \rhoSH(r_n, t_j) \, ,
\end{align}
where $\tau_{\scriptscriptstyle{\rm SH}} \sim \rsSH/\sigmaSH$ is the approximate timescale for heated particles to escape the subhalo.\footnote{The relevant timescale is not the local dynamical time, $\tau_{\rm dyn}(r) \propto(G\rhoSH(r))^{-1/2}$, but rather the orbital time, $\tau_{\rm orb}(r)\propto(G\bar{\rho}_{\scriptscriptstyle{\rm SH}}(r))^{-1/2}$, where $\bar{\rho}_{\scriptscriptstyle{\rm SH}}(r)$ is the average subhalo density enclosed in radius $r$. The latter is determined by the enclosed mass and is therefore less sensitive to the strongly depleted densities near the stripped halo boundary.}} Eq.~\eqref{eqn:rho_stripping} assumes that the instability saturates on a timescale shorter than the integration timestep, $\tau_{\rm sat} \lesssim 10^3 \omega_p^{-1} < \delta t$. Following saturation, turbulent heating broadens the subhalo velocity distribution, causing particles with velocities above the local escape speed to become unbound. The remaining bound material subsequently re-virializes on the subhalo orbital timescale $\tau_{\scriptscriptstyle{\rm SH}}$, leaving a bound mass fraction $f_{\rm bound}$.

The mass-loss model implemented here evolves the subhalo density profiles in a manner that is spherically symmetric. This assumption inevitably breaks down when the morphology of the subhalo is altered as a result of tidal effects at later times in its orbit. To avoid this, we only model the evolution of the subhalo up to the first pericentric passage, setting a cutoff time at $t = 1.2 \, t_{\rm peri}$. The fact that we track the subhalo only $\sim 1$~Gyr also mitigates the consequences of not modeling the growth of the host halo or disk as a function of time.

Lastly, we do not model dynamical friction acting on the subhalo's orbit and have verified that its inclusion only causes variations up to $1\%$ in the orbital path of the $10^7$ and $10^9~{\rm M}_\odot$ subhalos through first pericenter. Moreover, we neglect ram-pressure deceleration from dark Coulomb interactions between subhalo and MW particles. In the collisionless regime of interest, the mean free path for such interactions far exceeds the subhalo size, making the momentum transfer negligible. 

\subsection{Inputs from PIC Simulations} \label{sec:simulations}

We use the TRISTAN-MP~\citep{tristan-mp-pu} PIC code to model the dark plasma dynamics. 
While designed for the Standard Model, it can be applied to dark plasmas by exploiting the invariance of the Vlasov-Poisson equations under coordinate rescalings: $t \to \omega_{\rm p} t$, $v \to v/\sigma$, $x \to x/\lambda_{\scriptscriptstyle{\rm D}}$, $\Phi_{\rm g,em} \to \Phi_{\rm g,em}/\sigma^2$. Invariance under these transformations implies that the dynamics of a dark fermion and its antiparticle, characterized by a generic charge-to-mass ratio, can be faithfully modeled using simulations of $e^\pm$ plasmas, provided that the results are rescaled by the appropriate parameters.

We perform one-dimensional~(1D) PIC simulations using a uniform spatial grid with periodic boundary conditions. The grid discretizes physical space into $N_{\rm grid}$ cells of size $\Delta x$, on which the electromagnetic fields are defined, while particles are advanced continuously in phase space. The spatial resolution is set by $\Delta x$, while the velocity-space resolution is controlled by the number of particles per cell. We adopt $N_{\rm grid} = 2 \times 10^4$ and vary the particles-per-cell between $\sim 10$ and $2 \times 10^3$, with larger values required to resolve narrow velocity-space features associated with cold subhalos. We have confirmed that the results are unchanged when we double and half the number of grid points (see Figure~\ref{fig:1d_2d_comparison}).
The plasma Debye length of the subhalo, $\lambda_{\scriptscriptstyle{\rm D,SH}}$, is the smallest physical scale that must be resolved in the PIC simulation, requiring a grid spacing $\Delta x \lesssim \lambda_{\scriptscriptstyle{\rm D,SH}}$~\citep{1991ppcs.book.....B}. For the initial background plasma to be homogeneous over the simulation box, its total size ($N_{\rm grid} \Delta x$) must be much less than the subhalo's scale radius, giving
\begin{align}\label{eqn:param_constraint}
    {q_\chi \over m_\chi} \gg {N_{\rm grid} \over \sqrt{8 \pi} M_{\rm pl}} \, ,
\end{align}
where $M_{\rm pl} =1/\sqrt{8 \pi G}$ is the Planck mass.  For our simulations, this requires $q_{\chi} / m_{\chi} \gg 10^{-15} \ {\rm GeV}^{-1}$. The leading constraints coming from the Bullet Cluster require $q_\chi/m_\chi < 2 \times 10^{-14}\ \mathrm{GeV}^{-1}$~\citep{DeRocco:2024ifs}, which may suggest that the parameter space accessible to our semi-analytic model is narrow. However, these constraints are derived in the limit of a massless dark photon. For mediator masses in the range $\lambda_{\scriptscriptstyle{\rm D,BC}}^{-1} \lesssim m_{A'} \lesssim \lambda_{\scriptscriptstyle{\rm D,SH}}^{-1}$, where $\lambda_{\scriptscriptstyle{\rm D,BC}}$ and $\lambda_{\scriptscriptstyle{\rm D,SH}}$ are the Debye lengths in the Bullet Cluster and subhalo environments, respectively, plasma instabilities can operate efficiently in subhalos while remaining suppressed in the Bullet Cluster.\footnote{Since the dark photon in our minimal model does not kinetically mix with the Standard Model photon and need not constitute any fraction of the DM, the only relevant constraint on its mass arises from black hole superradiance~\citep{PhysRevD.103.095028, Cardoso:2018tly, PhysRevD.104.095029}. Spin measurements of stellar mass black holes exclude the range $6.5 \times 10^{-15} \, {\rm eV} < m_{A'} < 2.9 \times 10^{-11} \, {\rm eV}$, which lies above the fiducial masses considered in Eq.~\eqref{eqn:dark_photon_mass}.} In this regime, values of $q_\chi/m_\chi$ larger than $2 \times 10^{-14}\ \mathrm{GeV}^{-1}$ are allowed. At larger couplings, the strongest constraints arise from collisional effects. These include halo ellipticity measurements of NGC720~\citep{Agrawal:2016quu}, which constrain $q_\chi^4/m_\chi^3 \lesssim 8 \times 10^{-10} \, {\rm GeV}^{-3}$, and observations of the Bullet Cluster~\citep{Randall:2008ppe}, which constrain $q_\chi^4/m_\chi^3 \lesssim 2 \times 10^{-4} \, {\rm GeV}^{-3}$.

Importantly, current PIC codes do not include gravity. To apply them to our particular use case, we discretize the orbit of the subhalo and evaluate the instability growth rate for a specific location within the subhalo at an orbital time step.  This procedure is valid so long as the timescale for the instability to saturate is much smaller than the gravitational dynamical time.  As proven in  Appendix~\ref{app:gravitational_effects}, this is automatically satisfied in the parameter space set by Eq.~\eqref{eqn:param_constraint}.

A PIC simulation takes as input \emph{(i)}~the density ratio of the beams, which corresponds to the subhalo-to-MW DM density ratio at a specific location, and \emph{(ii)}~the beam speed, which corresponds to the subhalo's speed. The distribution function is evolved on a 1D grid, but because electric fields can have transverse components, which can introduce velocity components in transverse directions, two velocity dimensions are included. One spatial dimension is sufficient to capture longitudinal, electrostatic instabilities, such as the two-stream and beam-plasma instabilities, whose fastest-growing modes are aligned with the streaming direction. These modes dominate the early-time evolution of the distribution function. While higher-dimensional simulations allow transverse modes (e.g., Weibel/filamentation), the growth rates of these modes are parametrically suppressed by a factor of $\vSH$ \citep{Weibel:1959zz, Lasenby:2020rlf}. As a result, 1D simulations capture the leading-order dynamics and provide a first estimate of the efficiency of this process.  A more detailed discussion of this point is provided in Section~\ref{sec:discussion}. 

To account for the range of plasma conditions in each radial slice of the subhalo during its orbit, we carry out a suite of simulations that vary over the density ratio $\rhoSH/\rhoMW$ and subhalo speed $\vSH$. In particular, we consider ten roughly logarithmically-spaced density ratios in the range $\rhoSH/\rhoMW \in [10^{-3}, 10^{2}]$ and subhalo speeds  $\vSH/\sigmaMW = 1, 2, 3, 3.5, 4, 5$.  The 60 simulations provide an interpolation table for determining the bound fraction, $f_{\rm bound}$, for an arbitrary density ratio and subhalo speed. For each simulation, we define the bound fraction as the fraction of particles in the final distribution satisfying $|\vec{v} - \vec{v}_{\scriptscriptstyle{\rm SH}}| \le 4\sigmaSH$. Outside the simulated density range, we conservatively set the instability-induced mass loss to zero, corresponding to $f_{\rm bound}=1$, rather than extrapolating the simulation results into an untested regime. For $\rhoSH/\rhoMW > 10^{2}$, the large density contrast makes the simulations computationally expensive due to the large number of particles required to resolve the subhalo in velocity space. Furthermore, the instability growth rate is suppressed at large density ratios (see Eq.~\eqref{eqn:gamma_max}), providing additional physical motivation for this prescription. The resulting mass loss is dominated by modest density ratios, which are encountered near first pericenter for the inner shells and in the outskirts of the host halo for the outer shells.

Figure~\ref{fig:f_bound_arrows} shows representative grids of the bound mass fraction, $f_{\rm bound}$, as a function of the density ratio $\rhoSH/\rhoMW$ and subhalo velocity $\vSH/\sigmaMW$ for subhalos with masses of $10^7$ and $10^9~{\rm M}_\odot$. For both masses, stripping is most efficient for large subhalo velocities and moderate density ratios, $\rhoSH/\rhoMW \sim 1$, where plasma instabilities grow most rapidly. Finally, mass loss is more pronounced for the $10^7 \, {\rm M}_\odot$ subhalo because of its lower internal velocity dispersion: even relatively modest heating is sufficient to unbind a significant fraction of its DM.

Returning to the subhalo example highlighted in Figure~\ref{fig:subhalo_MW}, the right panel shows the evolution of the speed distribution $f(v)$ for a specific point in the orbiting subhalo before and after saturation of the instability.  As time progresses, resonant wave-particle interactions ``heat'' the cold subhalo population in an attempt to reduce velocity-space gradients, which manifests as a flattening of the distribution function between the peaks. Subhalo particles heated to speeds greater than the local escape velocity are stripped from the subhalo and mix with the background MW population.  

\section{Results} \label{sec:results}

The semi-analytic prescription developed here self-consistently couples the local, plasma instability-driven mass loss to the global orbital evolution, ensuring that the subhalo density profile evolves in concert with both its internal dynamics and its interaction with the MW environment. Next, we illustrate how instability-driven mass loss impacts the overall evolution of  a subhalo.  Section~\ref{sec:massloss} presents the mass-loss mechanisms at play in an individual subhalo, and Section~\ref{sec:shmf} discusses the ramifications for the subhalo population in a MW-mass host.

\subsection{Comparison of Mass Loss}
\label{sec:massloss}

This subsection compares the results for subhalos that only exhibit tidal stripping versus those that only exhibit plasma heating.  In reality, a subhalo orbiting a host in a dark plasma would experience both simultaneously, but we treat them separately to compare the relative magnitudes of the two effects. Crucially, for a dark force that is stronger than gravity, evaporation due to dark plasma instabilities occurs faster than the analogous effect in CDM---tidal stripping---leading to more dramatic effects.

\begin{figure}
    \centering
    \includegraphics[width=\linewidth]{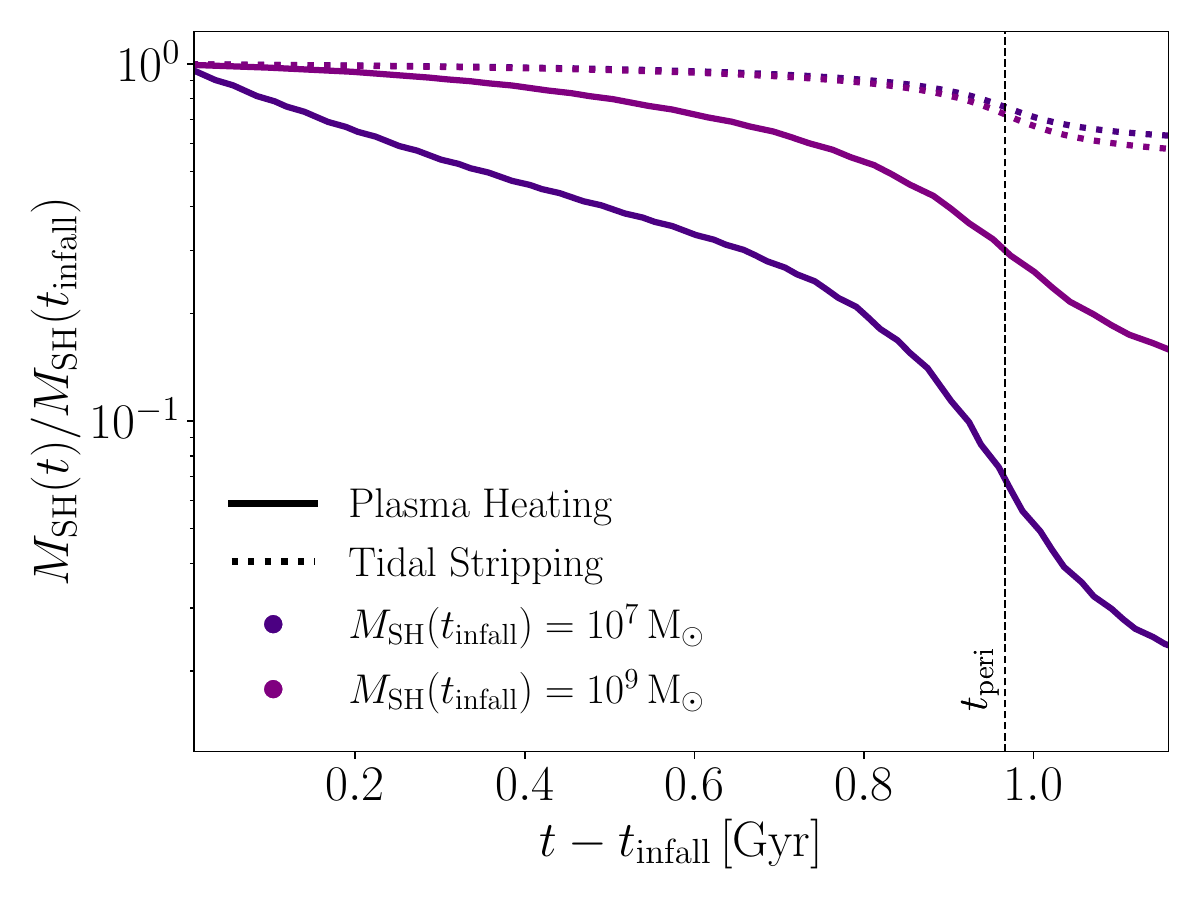}
    \caption{Subhalo mass evolution due to turbulent dark plasma heating and tidal stripping (both effects are treated in isolation of one another) for a charge-to-mass ratio $q_\chi/m_\chi = 10^{-14} \, e/m_p$ and an orbit with eccentricity $\epsilon = 0.9$. The fractional mass loss is plotted as a function of time for subhalos with infall virial mass $10^9 \ {\rm M}_\odot$~(violet) and $10^7 \ {\rm M}_\odot$~(indigo).
    The dotted lines denote the mass evolution from tidal stripping, while the solid lines denote the mass evolution from plasma heating.  The vertical black-dotted line denotes the time of first pericenter, $t_{\rm peri}$.}
    \label{fig:mass_loss_orb1}
\end{figure}

Figure~\ref{fig:mass_loss_orb1} shows the mass evolution for the $10^7$~(indigo) and $10^9 \, {\rm M}_\odot$~(violet) subhalo on an $\epsilon = 0.9$ orbit.  When the subhalo only exhibits tidal stripping~(dotted lines), {it loses $\sim 30\%$ of its infall mass by $t = 1.2 \, t_{\rm peri}$ for both virial masses considered.}
When the subhalo only exhibits plasma heating~(solid lines), the degree of mass loss is more accentuated: $\sim 84\%$ and $\sim 97\%$ of the initial mass is lost by $t = 1.2 \, t_{\rm peri}$ for the $10^9 \, \rm M_\odot$ and $10^7 \, \rm M_\odot$ subhalo, respectively. Because less-massive subhalos have colder DM velocity distributions and correspondingly lower escape velocities, even modest turbulent heating can unbind a large fraction of particles. 

\begin{figure*}
    \centering
    \includegraphics[width=\linewidth]{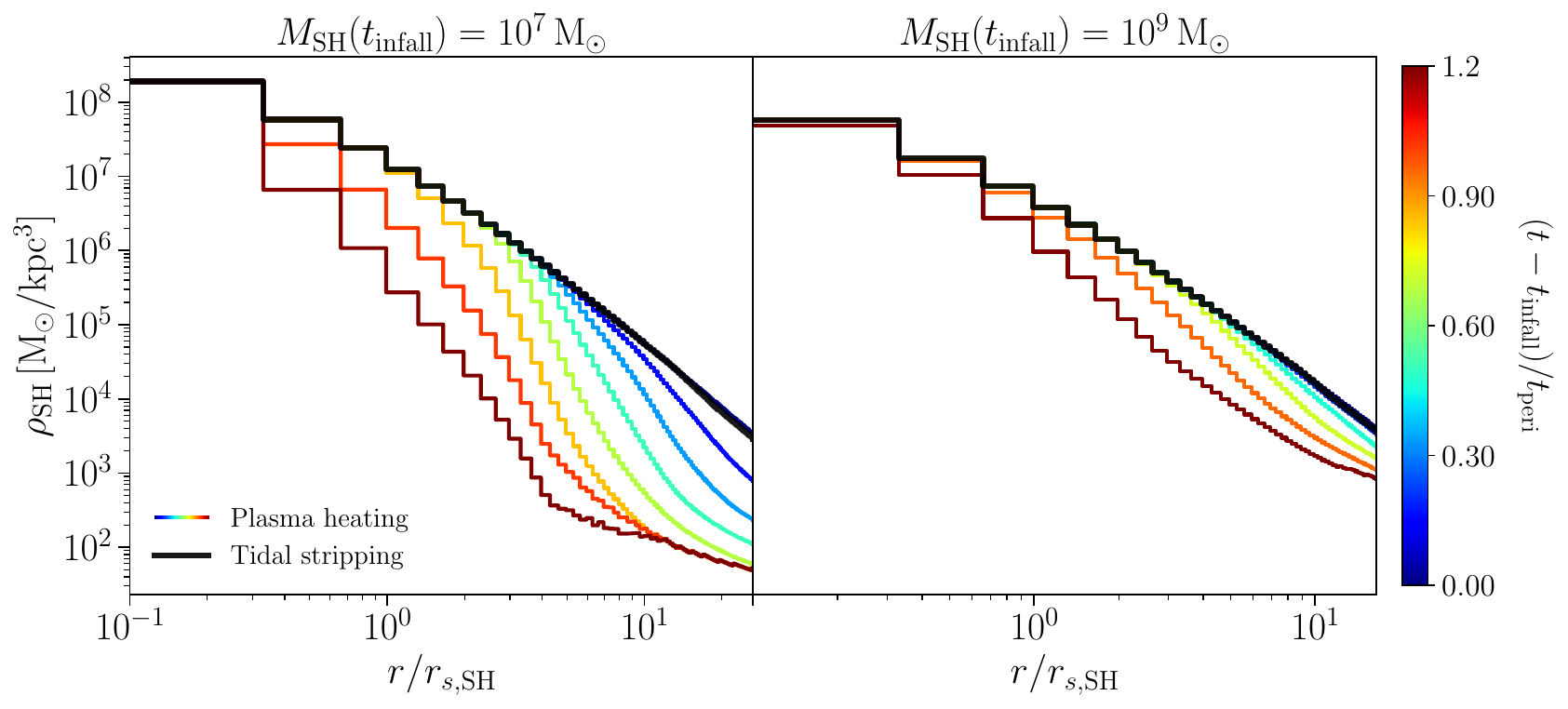}
    \caption{Density profile evolution for subhalos of infall virial mass $10^7 \ \rm {\rm M}_\odot$ (left panel) and $10^9 \ \rm {\rm M}_\odot$ (right panel) during an orbit with eccentricity $\epsilon = 0.9$. The charge-to-mass ratio is taken to be $q_\chi/m_\chi = 10^{-14} \, e/m_p$. The solid black line shows the final density profile at $t = 1.2 \, t_{\rm peri}$ when only tidal stripping is active.  Tidal stripping leaves the inner regions of both subhalos untouched, only removing mass outside the radius of truncation.  For comparison, the rainbow-colored lines demonstrate how the density profile evolves due to plasma heating, with blue~(red) indicating times near infall~(pericenter).  In this case, mass loss occurs in an ``outside-in'' fashion and does eventually affect the subhalo within its scale  radius, $\rsSH$, especially near pericenter.  Mass evaporation from plasma heating is a function of the subhalo density as well as the relative density of its dark plasma compared to that of the background halo. Local mass loss is only implemented for $\rhoSH/\rhoMW \in [10^{-3}, 10^2]$}
    \label{fig:density_profiles}
\end{figure*}

Figure~\ref{fig:density_profiles} illustrates the evolution of the  density profiles on an orbit with eccentricity $\epsilon = 0.9$, for the $10^7~{\rm M}_\odot$~(left panel) and $10^9~{\rm M}_\odot$~(right panel) subhalo. The solid black line corresponds to the final state at $t = 1.2 \, t_{\rm peri}$ of the density profile after undergoing tidal stripping. The resulting density profile agrees with the initial NFW up to a radius of truncation corresponding to the tidal radius of the subhalo. Beyond this tidal radius, the density profile is well described by either exponential or power-law suppression depending on the orbit, consistent with models in \citet{Hayashi_2003, Kazantzidis_2004, Pe_arrubia_2010}.

As shown in Figure~\ref{fig:density_profiles}, plasma heating dramatically affects the density profile evolution relative to tidal stripping.  As the subhalo travels towards first pericenter, its speed $\vSH$ increases, which enables instability growth over a larger fraction of its interior. For example, {in the case of the $10^9 \, \rm M_{\odot}$ subhalo}, an orbital speed close to $\vSH\sim 150$~km/s leads to $f_{\rm bound}\sim 0.8$ for a small range of density ratios around $\rhoSH/\rhoMW \sim 10^{-1}$.  In contrast, orbital velocities close to $\vSH \sim 400$~km/s lead to $f_{\rm bound} \sim 0.3$ for a density ratio $\rhoSH/\rhoMW \sim 1$ (see Figure~\ref{fig:f_bound_arrows}). Alongside this effect, the subhalo also moves through more dense regions of the background halo as it approaches first pericenter, which also provides more ideal conditions for instabilities to form near its center. Subhalo evaporation thus proceeds in an ``outside-in'' manner: the outer regions become susceptible to dark plasma instabilities earlier than the inner regions, leading to preferential heating and unbinding of particles in the outskirts before the core is affected. For both the  $10^7~{\rm M}_\odot$ and $10^9~{\rm M}_\odot$ subhalo with orbital eccentricity $\epsilon=0.9$, the density profile within the scale radius, $r<\rsSH$, is not significantly modified until $t = 1.1 \, t_{\rm peri}$. 

\begin{figure*}
    \centering
    \includegraphics[width=\linewidth]{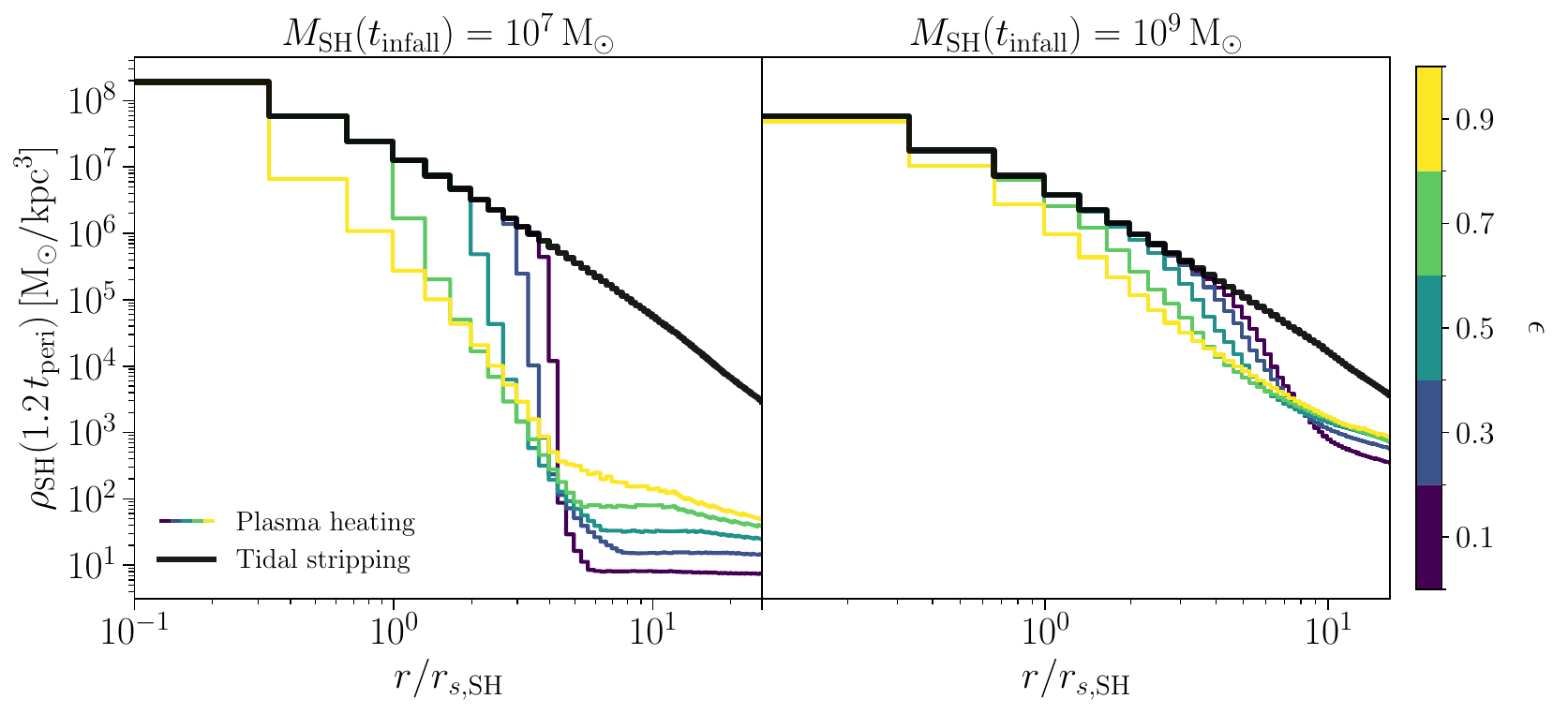}
    \caption{Density profiles near pericenter for subhalos with infall mass $10^7 \ {\rm M}_\odot$~(left panel) and $10^9 \ {\rm M}_\odot$~(right panel), as a function of eccentricity and for a charge-to-mass ratio $q_\chi/m_\chi = 10^{-14} \, e/m_p$. The solid black line, identical to that in Figure~\ref{fig:density_profiles}, corresponds to the density profile at $t=1.2 t_{\rm peri}$ in the $\epsilon = 0.9$ orbit when only tidal stripping is active. The colored lines denote the density profiles at $t=1.2 t_{\rm peri}$ of subhalos on orbits characterized by five different eccentricities.  Only subhalos on the most eccentric orbit have density profiles affected within $\rsSH$. As in Figure~\ref{fig:density_profiles}, the plasma heating prescription is applied only over the range $\rhoSH/\rhoMW \in [10^{-3}, 10^2]$.}
    \label{fig:density_profiles_first_pericenter}
\end{figure*}

Figure~\ref{fig:density_profiles_first_pericenter} shows how a subhalo's  density profile depends on orbital eccentricity at a fixed time of $1.2 \, t_{\rm peri}$ (note that $t_{\rm peri}$ is different for each of the orbits). For the highest eccentricity orbit ($\epsilon = 0.9$), the density profiles at first pericenter have mass loss occurring in the inner regions ($r \lesssim \rsSH$) of the profile. The less-eccentric orbits with $\epsilon = 0.1\text{--}0.5$ exhibit mass loss that is isolated to the outer regions ($r \gtrsim \rsSH$) of the subhalo. This behavior arises because lower-eccentricity orbits spend a larger fraction of time in the low-density outskirts of the host halo. Since plasma instabilities are most efficient for modest subhalo-to-host density contrasts, they preferentially heat particles in the outer regions of the subhalo.
    
As shown in Figure~\ref{fig:density_profiles_first_pericenter}, there is a radial scale that defines a sharp truncation in the density profiles at first pericenter for the $10^7\,\rm M_{\odot}$ subhalos. Under the assumptions of the model (namely, $\sigmaMW$ and $\sigmaSH$ being constant), at any point within an orbit, there exists a radius of stability  where $\rhoSH(r_{\rm stab})/\rhoMW$ is sufficiently large that the region enclosed by $r_{\rm stab}$ is impervious to the formation of plasma instabilities.

\subsection{Comments on Subhalo Mass Function}
\label{sec:shmf}

The strong dependence of plasma stripping on subhalo mass (or, equivalently, internal velocity dispersion) has direct implications for the MW subhalo mass function (SHMF). In particular, because plasma stripping is substantially more efficient in low-mass subhalos than tidal stripping, we expect the dark plasma scenario to preferentially suppress the SHMF for masses $\lesssim 10^7 \, {\rm M}_\odot$ relative to CDM. To estimate this effect, we compute the population-averaged mass loss for subhalos with initial (infall) masses of $10^7 \, {\rm M}_\odot$ and $10^9 \, {\rm M}_\odot$. We initialize each subhalo at infall with an orbital eccentricity drawn from the probability distribution~\citep{Jiang_2021}
\begin{align}
    \frac{dP}{d\epsilon} = \frac{\pi}{2} \frac{\epsilon \, \sin\left( \pi \sqrt{1 - \epsilon^2} \right)}{\sqrt{1 - \epsilon^2}} \, .
\end{align}
For subhalos with a fixed infall mass, $M_{\rm SH}(t_{\rm infall})$, we approximate the population-averaged final mass as
\begin{align}
    \bar{M}_f \approx \sum_{i = 0}^4 w_i \, M_{\rm SH}\!\left(1.2 \, t_{\rm peri}; \epsilon = 0.1(2i + 1)\right),
\end{align}
where $w_i = \int_{0.2 i}^{0.2 (i + 1)} d\epsilon \, \frac{dP}{d\epsilon}$, is the probability associated with the $i$th eccentricity bin, and $M_{\rm SH}(1.2\,t_{\rm peri}; \epsilon)$ is the subhalo mass at time $1.2 \, t_{\rm peri}$, computed in Section~\ref{sec:methodology}. We find $\bar{M}_f(10^7 \, {\rm M}_\odot) = 5\times 10^5 \, {\rm M}_\odot$ and $\bar{M}_f(10^9 \, {\rm M}_\odot) \approx 2 \times 10^8 \, {\rm M}_\odot$.

The preferential stripping of lower-mass subhalos produces potentially observable signatures in the SHMF. Because less-massive systems experience disproportionately high fractional depletion, size-dependent stripping shifts them to even lower final masses. This reduces the overall abundance of remnants at any given scale, suppressing the low-end of the SHMF and resulting in a shallower slope relative to CDM predictions. Additionally, if the stripping is extreme enough to completely unbind the halos, a sharp lower-bound cutoff may emerge. A quantitative prediction of this modified SHMF requires modeling the full population of subhalo orbits and their evolutionary histories, which is left to future work.

\section{Discussion} 
\label{sec:discussion}

The primary results of this paper were presented for electrostatic instabilities in one spatial dimension. One may wonder whether the qualitative features of the non-linear evolution of electrostatic streaming instabilities persist in higher spatial dimensions. Early two- and three-dimensional simulations reveal several qualitative differences from the 1D case. For instance, 1D simulations exhibit long-lived phase-space structures called Bernstein, Greene, Kruskal~(BGK) modes in the non-linear phase of streaming instabilities~\citep{PhysRev.108.546}. In higher dimensions, these structures dissipate more rapidly, depositing heat into the plasma~\citep{Morse1969}; this suggests that plasma heating and evaporation are more efficient in two- and three-dimensions. We confirm this expectation  using a 2D-2V simulation (two spatial and two velocity-space dimensions) initialized with density ratio $\rhoSH/\rhoMW = 1/2$, $\sigmaSH/\sigmaMW \approx 0.1$, and relative streaming speed $\vSH = 3 \sigmaMW$. The results, summarized in Figure~\ref{fig:1d_2d_comparison}, demonstrate that the bound fraction, $f_{\scriptscriptstyle{\rm bound}}$, at late times is lower in 2D than in 1D, consistent with heating and evaporation becoming more efficient in higher spatial dimensions. While further study is needed to extrapolate the 2D-2V results to other regions of parameter space, if the general trends hold, then the mass loss inferred from our 1D simulations is conservative.

In higher spatial dimensions, additional instabilities may also be excited. One important example is the Weibel, or ``filamentation,'' instability~\citep{1959PhFl....2..337F, Weibel:1959zz}, which arises from electromagnetic perturbations transverse to the streaming direction. In the cold limit, $\vSH \gg \sigmaSH,\sigmaMW$, its growth rate scales as $\gamma_{\scriptscriptstyle{\rm Weibel}} \sim \omega_{p,>} \vSH (\rho_</\rho_>)^{1/2}$, where $\rho_>$ and $\rho_<$ denote the densities of the denser and rarer streams, respectively, and $\omega_{p,>}$ is the plasma frequency of the denser stream~\citep{2010PhPl...17l0501B}. Relative to the electrostatic instability in the same limit, the Weibel growth rate is therefore parametrically suppressed by a factor of $\sim \vSH(\rho_</\rho_>)^{1/6}$. Nevertheless, the Weibel instability can saturate at a larger electromagnetic energy density and may therefore have an even more pronounced effect on subhalo evolution~\citep{Cruz:2022otv, DeRocco:2024ifs}. The resulting amplification of large-scale dark magnetic fields could also, in principle, shield subhalos from the surrounding MW dark plasma.

More generally, counter-streaming plasmas may be unstable to obliquely propagating electromagnetic modes, whose excitation is inherently multidimensional. Their growth rates can be comparable to that of the electrostatic instability and, for relativistic beams, may even exceed it~\citep{2010PhPl...17l0501B}. A detailed study of the nonlinear evolution of the Weibel and oblique instabilities, and of their implications for subhalo survival, is left to future work.

Finally, this paper focuses on symmetric charge species, analogous to a pair plasma. However, many well-motivated dark sector models, including the Twin Higgs model~\citep{PhysRevLett.96.231802}, more closely resemble the Standard Model. In such scenarios, DM may consist of charged species with highly asymmetric masses, analogous to electrons and ions in an ordinary plasma. These systems support a qualitatively different spectrum of streaming instabilities, including the Buneman~\citep{PhysRevLett.10.285} and ion-acoustic instabilities (see, e.g., \citet{Bhattacharjee2017}). Extending our analysis to such mass-asymmetric dark plasmas warrants further investigation.

\section{Conclusions} \label{sec:conclusion}

This paper presented the first study of the non-linear evolution of DM subhalos orbiting within the MW potential in hidden-$U(1)$ (dark plasma) models. Through interactions with the ambient MW dark plasma, subhalos whose velocities exceed the host's velocity dispersion can trigger streaming instabilities, such as the classic two-stream or beam-plasma instabilities. We investigated, for the first time, the full non-linear evolution of these instabilities, demonstrating the impact of turbulent heating on the evaporation of subhalos.

We developed a semi-analytic model that tracks a subhalo's mass loss from both tidal stripping and plasma heating at points along its orbit.  This model integrates the results of a suite of 1D plasma simulations generated with TRISTAN-MP to determine the fraction of DM evaporated at a specified region within the subhalo. We quantified the resulting evolution of the mass and density profiles of subhalos with two characteristic masses: $10^9 \ {\rm M}_\odot$, representing an ultrafaint dwarf, and $10^7 \ {\rm M}_\odot$, representing a non-luminous dark object. {Plasma heating substantially modifies the mass and density profiles of low-mass subhalos and can dominate over tidal stripping.  For example, by first pericenter, plasma heating can remove $\sim 97~(84)\%$ of the mass of a $10^7~(10^9)~{\rm M}_\odot$ subhalo on a highly eccentric $(\epsilon=0.9)$ orbit.  For comparison, tidal stripping alone would only remove $\sim 30\%$ of the mass for either of these cases.  These estimates are sensitive to the eccentricity of the orbit; for the case of a highly circular orbit ($\epsilon = 0.1$), plasma heating removes only $\sim 50~(30) \%$ of the mass of a $10^7~(10^9)~{\rm M}_\odot$ subhalo. 

Subhalo mass loss proceeds in an outside-in manner: at low subhalo velocities, instabilities develop only in the subhalo's outer regions. As the subhalo accelerates during infall, kinetic instabilities are triggered at progressively smaller radii.  Because instability growth is most efficient at modest density contrasts between the background and subhalo, as well as large subhalo velocities, instabilities develop efficiently deep within the subhalo during pericentric passage. In contrast, subhalos on more circular orbits reach lower velocities and remain at larger galactocentric radii, suppressing instability growth, particularly in their inner regions.

Our semi-analytic modeling demonstrates that dark plasma physics can significantly impact galaxy formation and evolution. This finding motivates more detailed numerical studies that self-consistently incorporate both gravity and plasma effects. Advanced frameworks will not only refine our current predictions and extend them to parameter spaces with lower charge-to-mass ratios, but also allow us to explore the simultaneous interplay between tidal stripping and plasma heating. Acting in concert, these mechanisms could accelerate subhalo dissolution; as tidal stripping removes mass and lowers the subhalo's velocity dispersion, the system becomes even more vulnerable to streaming instabilities. Finally, tracking the tidal debris from these dissolving subhalos will enable future studies to assess any resulting backreaction on the satellite. 

Our analysis suggests that plasma heating can have a population-level impact on the SHMF. Because plasma-driven stripping is more efficient for light, cold subhalos than for their more massive counterparts, it suppresses the abundance of low-mass subhalos. For our fiducial charge-to-mass ratio, $q_\chi/m_\chi = 10^{-14} \, e/m_p$, this suppression may manifest as a shallower low-mass slope of the SHMF or even a cutoff below $M_{\rm SH} \sim 10^7 \ {\rm M}_\odot$. Improved simulation frameworks will accelerate studies of halo interactions in dark plasma models.  This would enable extending predictions to full populations of subhalos, sharpening predictions for the SHMF and the diversity of rotation curves.  The ultimate goal of such efforts would be to compare with data from e.g., ELVES~\citep{2022ApJ...933...47C}, SPARC~\citep{Lelli2016}, the SAGA Survey~\citep{2024ApJ...976..117M}, and Merian~\citep{Danieli2025}.  The results of this work indicate that dark plasma effects could indeed lead to potentially distinctive signatures when studying the mass functions and/or radial distributions of satellites in MW-mass hosts providing a critical avenue for constraining or discovering such models.

\clearpage

\section{Acknowledgments}

The authors would like to acknowledge helpful conversations and feedback from Robert Ewart, Dylan Folsom, Pierce Giffin, Jimmy Juno, Eliot Quataert, and Anatoly Spitkovsky. ML and AP are supported by the Department of Energy~(DOE) under Award Number DE-SC0007968. ML is also supported by the Simons Investigator in Physics Award. AP also acknowledges support from the Leinweber Foundation. The simulations and analyses presented in this article were performed on computational resources managed and supported by Princeton University’s Research Computing.

This report was prepared as an account of work sponsored by an agency of the United States Government. Neither the United States Government nor any agency thereof, nor any of their employees, makes any warranty, express or implied, or assumes any legal liability or responsibility for the accuracy, completeness, or usefulness of any information, apparatus, product, or process disclosed, or represents that its use would not infringe privately owned rights. Reference herein to any specific commercial product, process, or service by trade name, trademark, manufacturer, or otherwise does not necessarily constitute or imply its endorsement, recommendation, or favoring by the United States Government or any agency thereof. The views and opinions of authors expressed herein do not necessarily state or reflect those of the United States Government or any agency thereof.

\software{} This research made extensive use of the publicly available codes \texttt{IPython}~\citep{PER-GRA:2007}, \texttt{matplotlib}~\citep{Hunter:2007},
\texttt{Jupyter}~\citep{Kluyver2016jupyter},
\texttt{NumPy}~\citep{harris2020array}, and 
\texttt{SciPy}~\citep{2020SciPy-NMeth}. The simulation setup and analysis script used in this work are publicly available~\citep{GithubRepo}.

\bibliography{darkplasmas}

\clearpage

\onecolumngrid

\begin{center}
  \textbf{\large Supplementary Materials}\\[.2cm]
  \vspace{0.05in}
  {}
\end{center}

\vspace{0.2in}
\setcounter{equation}{0}
\setcounter{figure}{0}
\setcounter{table}{0}
\setcounter{section}{0}
\setcounter{page}{1}
\thispagestyle{empty}
\makeatletter
\renewcommand{\theequation}{S\arabic{equation}}
\renewcommand{\theHequation}{S\arabic{equation}}
\renewcommand{\thesection}{S\arabic{section}}
\renewcommand{\theHsection}{S\arabic{section}}
\renewcommand{\thefigure}{S\arabic{figure}}
\renewcommand{\thetable}{S\arabic{table}}
\def\set@footnotewidth{\onecolumngrid}

\vspace{-0.4in}

\section{The Subhalo Beam-Plasma System} \label{app:linear_stability}

This appendix reviews the linear and nonlinear stability of the equilibrium Milky Way~(MW)–subhalo~(SH) system described by Eq.~\eqref{eq:equilibriumf}. Plasma instabilities arise when the distribution function contains a source of free energy, such as a velocity-space gradient. In particular, counter-streaming populations can produce multiple peaks (or ``bumps'') in the velocity distribution, rendering the system susceptible to streaming instabilities. These instabilities extract free energy from the distribution function and transfer it to electromagnetic fluctuations. This work focuses on dark electrostatic streaming instabilities analogous to the familiar two-stream and beam-plasma instabilities (see, e.g.,~\cite{Bhattacharjee2017}).  App.~\ref{app:instability_growth} describes the growth rate for such instabilities, neglecting gravitational effects; the validity of this approximation is justified in App.~\ref{app:gravitational_effects}.

\subsection{Estimating the Instability Growth Rate}
\label{app:instability_growth}

The standard approach to studying the linear stability of a distribution function is to derive the dispersion relation for small perturbations to the distribution function and electrostatic potential. At a given point along a subhalo's orbit---and on scales much smaller than its scale radius---the equilibrium distribution may be treated as locally homogeneous and written as
\begin{align}
    f(\vec{v}) = 2 f_\pm(\vecv) = \frac{\rhoMW}{m_\chi (\pi \sigmaMW^2)^{3/2}} e^{- v^2/\sigmaMW^2} + \frac{\rhoSH}{m_\chi (\pi \sigmaSH^2)^{3/2}} e^{- (\vecv - \vecvSH)^2/\sigmaSH^2} \, ,
\end{align}
where $m_\chi$ is the dark matter mass, $\rho_{\scriptscriptstyle{\rm MW (SH)}}$ and $\sigma_{\scriptscriptstyle{\rm MW (SH)}}$ denote the local MW~(subhalo) density and velocity dispersion, respectively, and $\vecvSH$ is the instantaneous subhalo streaming velocity in the MW's rest frame. 

To study the evolution of small perturbations, we linearize the Vlasov-Poisson system about this equilibrium and consider perturbations of the form
\begin{align}\label{eqn:fpm_uniform}
f_\pm(\vecv) = \frac{f_{0}(\vecv)}{2} + \delta f_\pm e^{i k z - i \omega t} \qquad \text{and} \qquad 
\Phi_{\rm em} = \delta \Phi_{\rm em} e^{i k z - i \omega t},
\end{align}
where $\hat{z}$ is chosen to lie along the direction of the subhalo streaming velocity. Substituting Eq.~\eqref{eqn:fpm_uniform} into the linearized Vlasov-Poisson system yields the dispersion relation
\begin{align} \label{eqn:dispersion_relation_clean}
1 - \sum_{j \in \{\rm MW,SH\}}
\frac{q_\chi^2 \rho_j}{m_\chi^2 k^2 \sigma_j^2}
\left(1 + \zeta_j Z(\zeta_j)\right)
= 0 \, ,
\,\, \text{where} \,\,
\zeta_j = \frac{1}{\sqrt{2} \sigma_j}\left(\frac{\omega}{k} - v_j\right) \,,
\end{align}
$v_j$ denotes the bulk speed of stream $j$, and
$Z(\zeta) \equiv \pi^{-1/2}\int_{-\infty}^\infty dz\, e^{-z^2}/(z-\zeta)$ is the Plasma Dispersion Function. The imaginary part of $Z(\zeta)$ describes resonant wave-particle interactions and allows for both Landau damping and instability. Solutions of Eq.~\eqref{eqn:dispersion_relation_clean} with ${\rm Im}(\omega)>0$ correspond to exponentially growing modes. Physically, these unstable modes extract free energy from the relative streaming motion between the MW and SH populations and transfer it to collective electrostatic fluctuations.

In full generality, Eq.~\eqref{eqn:dispersion_relation_clean} is difficult to solve analytically. A useful limit that admits analytic treatment is the cold-beam regime, in which $\vSH \gg \sigmaMW, \sigmaSH$. This regime can be relevant during the pericentric passage of a subhalo on an eccentric orbit. In this limit, Eq.~\eqref{eqn:dispersion_relation_clean} reduces to
\begin{align} \label{eqn:dispersion_relation_cold_clean}
1 - \frac{\omega_{{\rm p},\scriptscriptstyle{\rm MW}}^2}{\omega^2} - \frac{\omega_{{\rm p},\scriptscriptstyle{\rm SH}}^2}{(\omega - \vSH k)^2} = 0 \,, \,\, \text{where} \,\, \omega^2_{p,\scriptscriptstyle{\rm MW (SH)}} \equiv \frac{q_\chi^2 \rho_{\scriptscriptstyle{\rm MW (SH)}}}{m_\chi^2} \,.
\end{align}
When $\rhoMW \sim \rhoSH$, the system approaches the symmetric two-stream limit, in which the fastest-growing mode has a growth rate $\gamma_{\rm max} \sim \omega_{\rm p}$, where $\omega_{\rm p}$ is the plasma frequency of either population. In the opposite limit, where one population is much denser than the other, the instability reduces to the beam-plasma instability, and the growth rate of the fastest-growing mode scales as
\begin{align} \label{eqn:gamma_max}
\gamma_{\rm max} \sim \omega_{{\rm p},>} \left(\frac{\omega_{{\rm p},<}}{\omega_{{\rm p},>}}\right)^{2/3} = \omega_{{\rm p},>} \left( \frac{\rho_{<}}{\rho_{>}} \right)^{1/3} \,,
\end{align}
where $\rho_{>(<)}$ and $\omega_{{\rm p},>(<)}$ denote the density and plasma frequency of the denser (rarer) population, respectively~\citep{1973ppp..book.....K}. Eq.~\eqref{eqn:gamma_max} shows that, in the cold limit, plasma instabilities are most efficient for modest density contrasts and become increasingly suppressed as the density contrast grows. As discussed in the main text, this suppression, together with the computational cost of simulating extreme density contrasts, motivates restricting our analysis to $\rhoSH/\rhoMW < 10^{2}$.

\subsection{Justification of Local Plasma Approximation}
\label{app:gravitational_effects}

This work models the subhalo-MW interaction using local plasma simulations that neglect gravity. Here, we justify this approximation and identify the region of parameter space in which it is valid. As described in Eq.~\eqref{eqn:fpm_uniform}, we consider an initial equilibrium configuration that is locally neutral, $f_\pm = f_0/2$, where $f_0$ is a solution to the collisionless Boltzmann equation in the absence of dark electromagnetism.

Linearizing about this equilibrium and defining the density and charge perturbations 
\begin{equation}
\delta f_\rho \equiv \delta f_+ + \delta f_- \quad \text{and} \quad \delta f_{\rm q} \equiv \delta f_+ - \delta f_- \, ,
\end{equation}
respectively, one finds that the linearized equations separate into two independent branches. The density perturbation~($\delta f_\rho$) couples only to the gravitational potential, while the charge perturbation ($\delta f_{\rm q}$) couples only to the electrostatic potential. Therefore, the electrostatic streaming instabilities studied in this work correspond to charge fluctuations and are not directly sourced by self-gravity at linear order.

Nevertheless, gravity enters through the slow spatial variation of the equilibrium distribution function. To estimate its importance, consider a halo (which could represent either the MW or subhalo) with characteristic scale radius~($r_{\rm s}$), velocity dispersion ($\sigma$), and density ($\rho$). The corresponding gravitational dynamical time is $t_{\rm dyn}  \sim {1}/{\sqrt{G\rho}}$, while electrostatic fluctuations evolve on the plasma timescale $\omega_{\rm p}^{-1} = {m_\chi}/({q_\chi \sqrt{\rho}})$. The ratio of these timescales is $\omega_{\rm p} t_{\rm dyn}\sim {q_\chi}/({m_\chi \sqrt{G}}) = \sqrt{8\pi} {q_\chi M_{\rm pl}}/{m_\chi}$. For the parameter space of interest, this ratio is typically very large
\begin{equation}
    \omega_{\rm p} t_{\rm dyn} \approx 4 \times 10^4 \left( \frac{q_\chi/m_\chi}{10^{-14} \, e/m_p} \right) 
\end{equation}
and also independent of density.  This implies that electrostatic instabilities grow much more rapidly than gravity can modify the halo.

A second requirement is that the characteristic wavelength of the most unstable modes be much smaller than the scale over which the halo varies, $r_{\scriptscriptstyle{s}}$. The relevant plasma scale is the Debye length, $\lambda_{\scriptscriptstyle{\rm D}}=\sigma/\omega_{\rm p}$.  The local approximation requires $\lambda_{\scriptscriptstyle{\rm D}} \ll r_{\scriptscriptstyle{s}}$, which is indeed satisfied---see Eq.~\eqref{eq:debye}. In this limit, the assumption of an approximately homogeneous ambient plasma is well-justified. 

Finally, since the nonlinear evolution of the instability is controlled by its saturation time, $\tau_{\rm sat} = N \gamma_{\rm max}^{-1}$ (where $N$ is the number of e-foldings required for saturation and $\gamma_{\rm max}$ is the rate of the fastest-growing mode), the most conservative criterion for neglecting gravity is $\tau_{\rm sat} \ll t_{\rm dyn}$. The smallest density ratio we simulate is $\rho_</\rho_> \gtrsim 10^{-3}$, therefore from Eq.~\eqref{eqn:gamma_max}, $\gamma_{\rm max}/\omega_{{\rm p},>} \equiv \tilde{\gamma}_{\rm max} \gtrsim 10^{-1}$. The condition that the saturation time be much less than the local dynamical time becomes
\begin{align}
    \frac{q_\chi}{m_\chi} \gg 2 \times 10^{-16} \frac{e}{m_p} \, \left( \frac{N}{100} \right) \left( \frac{\tilde{\gamma}_{\rm max}}{10^{-1}} \right)^{-1}, 
\end{align}
which is safely satisfied by the fiducial charge-to-mass ratio, $q_\chi/m_\chi = 10^{-14} \, e/m_p$, considered throughout the main text.

\clearpage
\section{Supplementary Figures}
\label{app:suppfigures}

This appendix provides some additional figures that supplement the discussion in the main text.
\vspace{0.2in}

\begin{figure*}[h]
    \centering
    \includegraphics[width=0.9\linewidth]{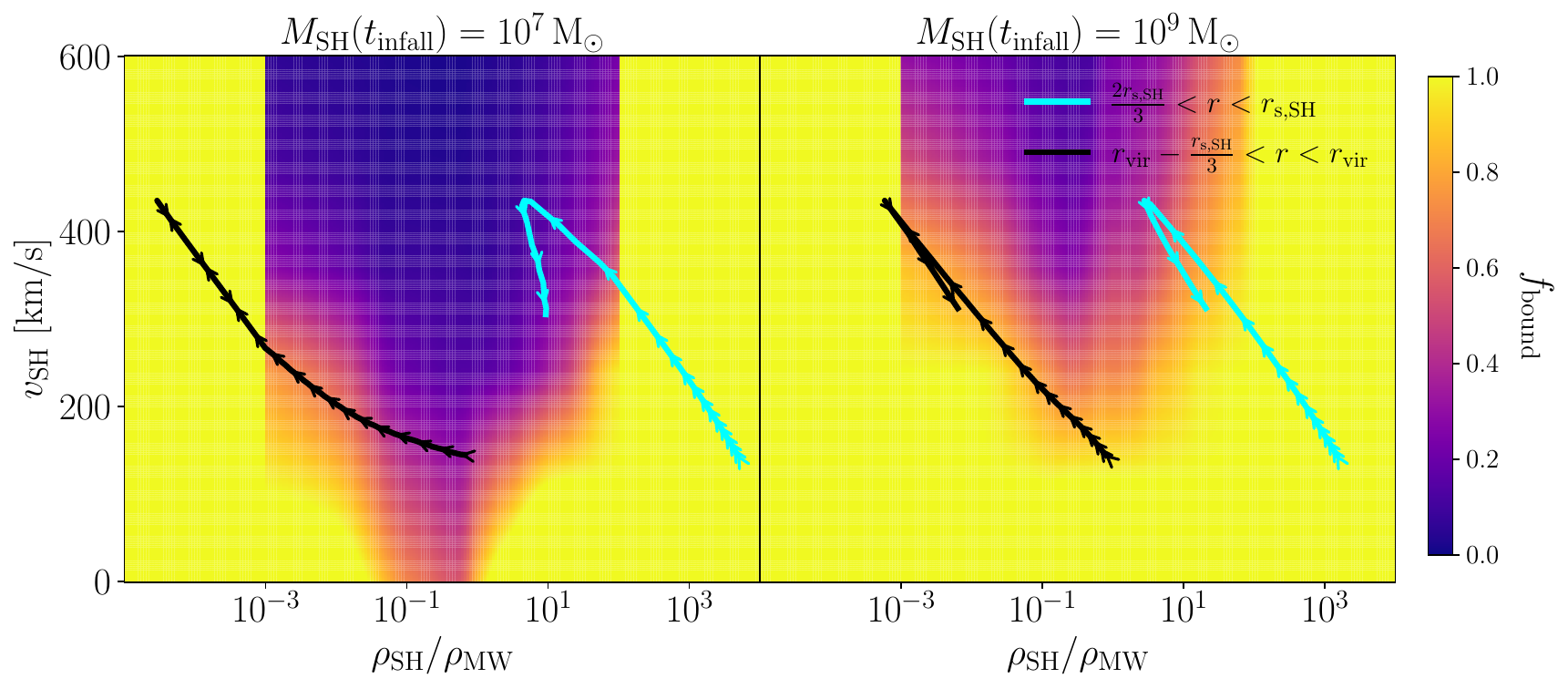}
     \caption{Interpolation of the bound fraction of particles after non-linear saturation of the beam–plasma instability, shown as a function of the density ratio, $\rho_{\rm SH}/\rho_{\rm MW}$, and the subhalo streaming velocity, $\vSH$, for subhalos of mass $M = 10^7 \, {\rm M}_\odot$ and $10^9 \, {\rm M}_\odot$ in the left and right panel, respectively. Also shown are the trajectories of an inner shell ($2 r_{\rm s, SH}/3 < r < r_{\rm s, SH}$, blue) and the outermost shell ($r_{\rm vir} - r_{\rm s, SH}/3 < r < r_{\rm vir}$, black) of each subhalo as they evolve along a high-eccentricity ($\epsilon = 0.9$) orbit from $t = 0$ to $t = 1.2 \, t_{\rm peri}$, with arrows moving along the direction of time. Mass loss is truncated at density ratios below $10^{-3}$ and above $10^2$, as this regime is not simulated. The figure shows that mass loss is most efficient at modest density ratios. For the inner and outer shells, this regime is encountered near the first pericenter and in the outskirts of the host halo, respectively.}
    \label{fig:f_bound_arrows}
\end{figure*}
\vspace{0.2in}

\begin{figure}[h]
    \centering
    \includegraphics[width=0.9\linewidth]{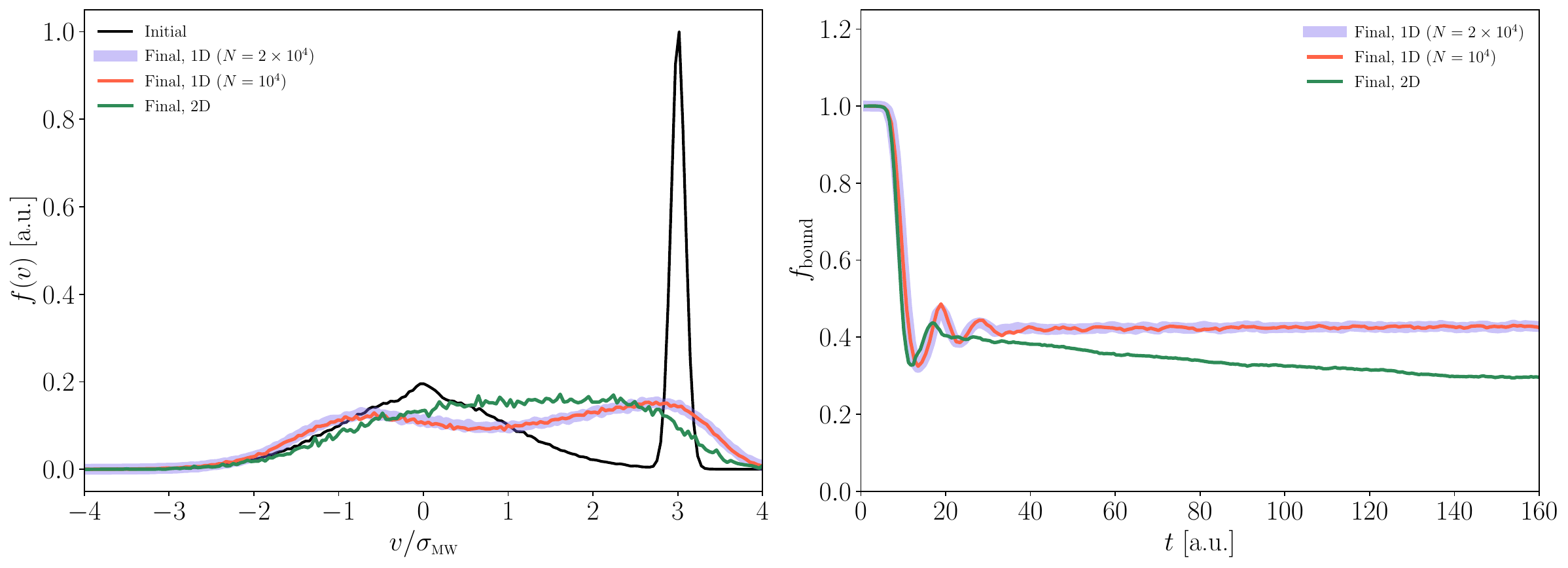}
    \caption{Comparison of the results of 1D-2V simulations with $N_{\rm grid} = 2 \times 10^4$ (thick purple) and $N_{\rm grid} = 10^4$ (orange) with those of a 2D-2V simulation (green), all performed using TRISTAN-MP. The fiducial simulations adopt $\rhoSH/\rhoMW = 1/2$ and $\sigmaSH/\sigmaMW \approx 0.1$, corresponding to a $10^9$~M$_\odot$ subhalo. The 2D-2V simulation uses an $N_x \times N_y = 1200 \times 800$ grid. The left panel compares the evolution of the particle distribution functions, with the initial distribution shown in black, while the right panel compares the bound fraction. At late times, the bound fraction asymptotes to a lower value in 2D than in 1D, indicating that stripping is more efficient in higher spatial dimensions, as discussed in Section~\ref{sec:discussion}.}
    \label{fig:1d_2d_comparison}
\end{figure}

\end{document}